\documentclass[fleqn,10pt]{wlscirep}
\usepackage[utf8]{inputenc}
\usepackage[T1]{fontenc}
\usepackage{lineno}
\usepackage{multirow}
\usepackage{multicol}
\usepackage{xcolor}%
\usepackage{colortbl}
\usepackage{epsfig}

\newcommand{\multirowcell}[3]{%
  \multirow{#1}{*}{\cellcolor{#2}#3}%
  \foreach \n in {1,...,#1} {
    \ifnum \n>1 \\[-\dimexpr\baselineskip-\arrayrulewidth\relax]
    \cellcolor{#2} \fi
  }%
}

\title{AI.vs.Clinician: Unveiling Intricate Interactions Between AI and Clinicians through an Open-Access Database}

\author[1,3,4,$\dag$]{Wanling Gao}
\author[5,$\dag$]{Yuan Liu}
\author[2,18]{Zhuoming Yu}
\author[1]{Dandan Cui}
\author[5]{Wenjing Liu}
\author[2,18]{Xiaoshuang Liang}
\author[2,18]{Jiahui Zhao}
\author[2,18]{Jiyue Xie}
\author[2,18]{Hao Li}
\author[5,9]{Li Ma}
\author[6]{Ning Ye}
\author[6]{Yumiao Kang}
\author[7]{Dingfeng Luo}
\author[8]{Peng Pan}
\author[10]{Wei Huang}
\author[11]{Zhongmou Liu}
\author[12]{Jizhong Hu}
\author[13]{Fan Huang}
\author[14]{Gangyuan Zhao}
\author[15]{Chongrong Jiang}
\author[16]{Tianyi Wei}
\author[17,*]{Zhifei Zhang}
\author[2,18,*]{Yunyou Huang}
\author[1,3,4,*]{Jianfeng Zhan}
\affil[1]{Institute of Computing Technology, Chinese Academy of Sciences, Beijing, 100190, China}
\affil[2]{Guangxi Key Lab of Multi-Source Information Mining and Security, Guangxi
Normal University, Guilin, 541004, China}
\affil[3]{International Open Benchmark Council}
\affil[4]{University of Chinese Academy of Sciences, Beijing, 100086, China}
\affil[5]{Guilin Medical University, Guilin, 541100, China}
\affil[6]{Affiliated Hospital of Guilin Medical University, Guilin, 541000, China}
\affil[7]{Xing An County People's Hospital, Guilin, 541300, China}
\affil[8]{Meng Shan County People's Hospital, Wuzhou, 543000, China}
\affil[9]{XuanJi Technology Co., Ltd., Guilin, 541000, China}
\affil[10]{Guilin People's Hospital, Guilin, 541000, China}
\affil[11]{Yong Fu County People's Hospital, Guilin, 541000, China}
\affil[12]{Ling Chuan County People's Hospital, Guilin, 541000, China}
\affil[13]{The Second Affiliated Hospital of Guilin Medical University, Guilin, 541000, China}
\affil[14]{Quan Zhou County People's Hospital, Guilin, 541000, China}
\affil[15]{Guan Yang County People's Hospital, Guilin, 541000, China}
\affil[16]{International College, Guangxi University, Nanning, 530004, China}
\affil[17]{Capital Medical University, Beijing, 100069, China}
\affil[18]{Key Lab of Education Blockchain and Intelligent Technology, Ministry of
Education, Guangxi Normal University, Guilin, 541004, China}

\affil[$\dag$]{co-first author}
\affil[*]{corresponding author(s): Jianfeng Zhan (zhanjianfeng@ict.ac.cn) and Yunyou Huang (huangyunyou@gxnu.edu.cn) and Zhifei Zhang (zhifeiz@ccmu.edu.cn)}

\begin{abstract}
Artificial Intelligence (AI) plays a crucial role in medical field and has the potential to revolutionize healthcare practices. 
However, the success of AI models and their impacts hinge on the synergy between AI and medical specialists, with clinicians assuming a dominant role.
Unfortunately, the intricate dynamics and interactions between AI and clinicians remain undiscovered and thus hinder AI from being translated into medical practice. 
To address this gap, we have curated a groundbreaking database called AI.vs.Clinician. 
This database is the first of its kind for studying the interactions between AI and clinicians. 
It derives from
7,500 collaborative diagnosis records on a life-threatening medical emergency -- Sepsis -- from 14 medical centers across China.
For the patient cohorts well-chosen from MIMIC databases, the AI-related information comprises the model property, feature input, diagnosis decision, 
and inferred probabilities of sepsis onset presently and within next three hours.
The clinician-related information includes the viewed examination data and sequence, viewed time,  preliminary and final diagnosis decisions with or without AI assistance, and recommended treatment.
\end{abstract}
\begin{document}

\begin{titlepage} 

	\centering 
	
	\scshape 
	
	\vspace*{\baselineskip} 
	
	
	\rule{\textwidth}{1.6pt}\vspace*{-\baselineskip}\vspace*{2pt} 
	\rule{\textwidth}{0.4pt} 
	
	\vspace{0.75\baselineskip} 
	
	{\LARGE AI.vs.Clinician: Unveiling Intricate Interactions Between AI and Clinicians through an Open-Access Database} 
	
	\vspace{0.75\baselineskip} 
	
	\rule{\textwidth}{0.4pt}\vspace*{-\baselineskip}\vspace{3.2pt} 
	\rule{\textwidth}{1.6pt} 
	
	\vspace{2\baselineskip} 
	
	
	
	\vspace*{3\baselineskip} 
	
	
	 \begin{center}Edited By\end{center}
	
	\vspace{0.5\baselineskip} 
	
	{ \begin{center} Wanling Gao \\ Yuan Liu \\ Zhuoming Yu \\ Dandan Cui \\ Wenjing Liu \\ Xiaoshuang Liang \\ Jiahui Zhao \\ Jiyue Xie \\ Hao Li \\ Li Ma \\ Ning Ye \\ Yumiao Kang \\ Dingfeng Luo \\ Peng Pan \\ Wei Huang \\ Zhongmou Liu \\ Jizhong Hu \\ Fan Huang \\ Gangyuan Zhao \\ Chongrong Jiang \\ Tianyi Wei \\ Zhifei Zhang \\ Yunyou Huang \\ Jianfeng Zhan \end{center}}
	
	
	\vspace{0.5\baselineskip} 

	\vfill 
	
	
	\epsfig{file=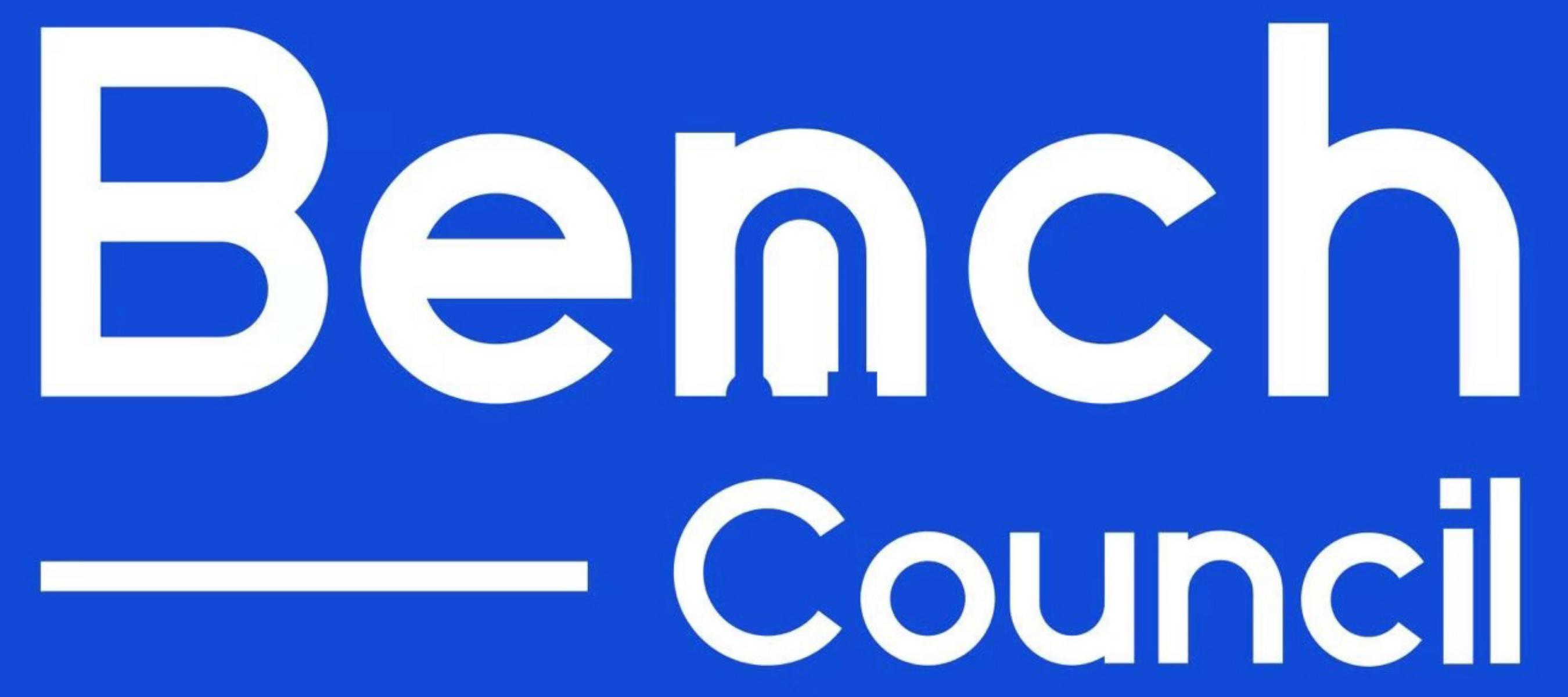,height=2cm}
	\textit{\\BenchCouncil: International Open Benchmark Council\\http://www.benchcouncil.org} 
	\vspace{5\baselineskip} 

	Technical Report No. BenchCouncil-AI.vs.Clinician-2024 
	
	{\large June 11, 2024} 

\end{titlepage}

\flushbottom
\maketitle

\thispagestyle{empty}


\section*{Background \& Summary}

Artificial Intelligence (AI), as a revolutionary technology, has attracted extensive research on its applications to medicine, leading to the emergence of a new research field known as AI in medicine~\cite{holmes2004artificial,rajpurkar2022ai,hamet2017artificial,topol2019high,he2019practical}.
Despite this advance, 
integrating AI models into real-world medical systems and clinical practice is still far away. One significant reason is that clinicians still play important roles~\cite{yun2021behavioral} and reflect unknown meanwhile unpredictable interactions with AI~\cite{rajpurkar2022ai}.
Meanwhile, human-in-the-loop is considered an essential developmental trend of AI in medicine and is gradually gaining attention~\cite{cohen2023ai,patel2019human,fosch2021human,yu2023pi,walsh2018human}.
However, the lack of open, accessible interaction data between AI and clinicians poses significant challenges to related research. It hinders AI improvements that can better assist clinicians in clinical practice.

Collaborated with 14 medical centers and 125 clinicians from December 2022 to May 2024, we perform clinical trials and release the AI.vs.Clinician database as the first human-AI interaction data adopting Sepsis diagnosis for the first step. On one hand, Sepsis holds immense significance in the field of medicine. It affects millions of people worldwide and is a leading cause of morbidity and mortality~\cite{mayr2014epidemiology,rudd2020global}. Understanding the complex mechanisms underlying Sepsis is crucial for early detection, accurate diagnosis, and effective treatment strategies.
On the other hand, such analysis is also applicable and valuable for interaction analysis in other medical issues.

AI.vs.Clinician is an extensive and human-centered database that comprises information related to the behavior variations of clinicians' diagnoses with or without the assistance of different AI models. The database contains the data records of patient cohorts, AI models, and clinicians. In terms of patient cohort, the database contains 3000 patient cases with current and history examination data to be diagnosed by AI models and clinicians. 
From the perspective of AI models, the database provides four models with different model types (i.e., traditional machine learning, deep learning) and properties (i.e., quality type, quality number), feature input for training, and their inference results on every patient case including probabilities of sepsis onset presently and within next three hour. As a control, we also introduce a random model in our clinical settings. Note that the 3-hour window aligns with the treatment guidelines from the CMS sepsis core measure (SEP1) and the Surviving Sepsis Campaign~\cite{henry2022factors,singer2016third,levy2018surviving}.
The clinician-related information records all their operations and corresponding timestamps for preliminary diagnosis and treatment and final diagnosis and treatment.

We expect that AI.vs.Clinician will have broad international usage in various domains, including academic and industrial research, comprehension of the complex interaction, quality improvement of AI models, and further facilitate the advancement and practical implementation in AI in medicine. 


\begin{figure}[htbp]
\centering
\includegraphics[scale=0.38]{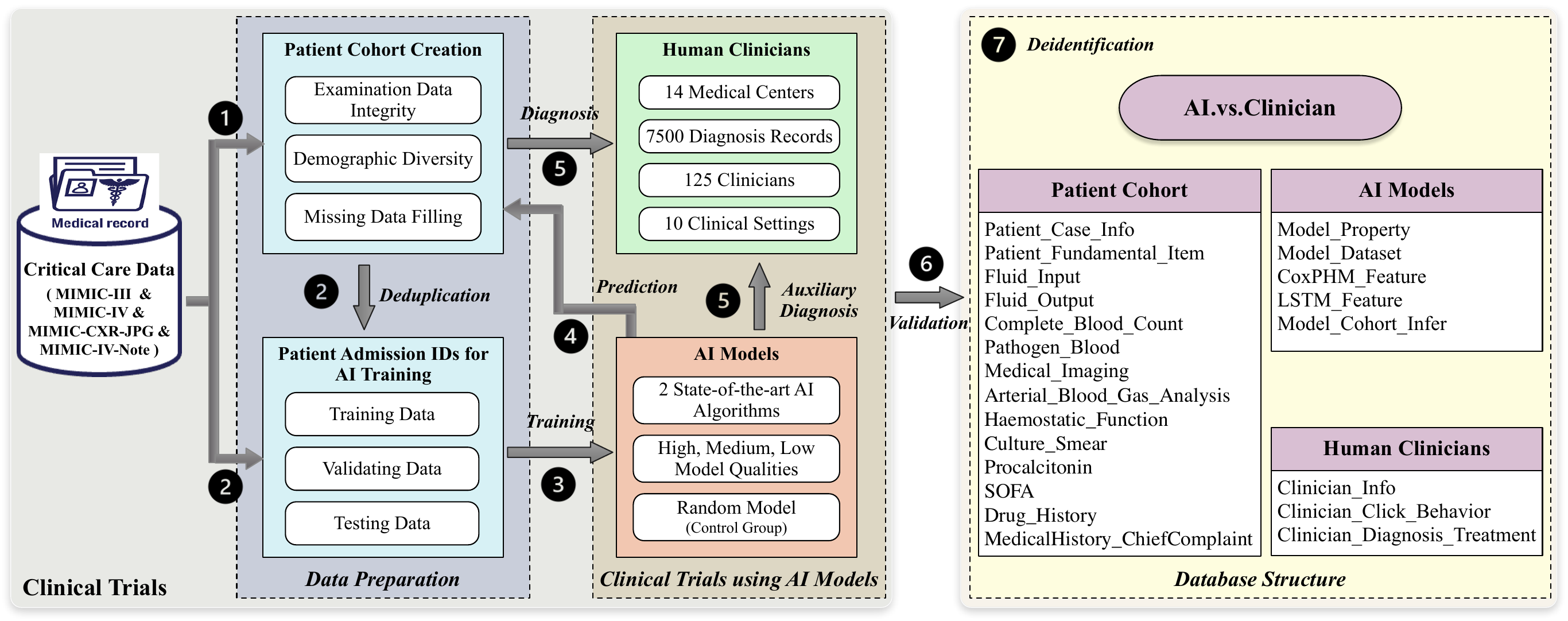}
\caption{\textbf{AI.vs.Clinician Database Development Process.} The human-AI interaction data are acquired from the clinical trials using AI models in 14 medical centers. A patient cohort is created based on the MIMIC databases~\cite{johnson2016mimic,johnson2020mimic,johnson2019mimic,johnson2023mimic} and used for the collaborative decision-making of AI models and clinicians. Finally, the collected data are validated and de-identified. Tables in AI.vs.Clinician are classified into three categories: patient cohort that records the patients' information and examination data; AI models that record the AI model related information and inference results on patient cohort, and clinicians that record the clinician information and interaction behaviors with AI models. }
\label{doctor}
\end{figure}



\section*{Methods}

\textbf{Ethics statement.} 
This research has been approved by the Ethics Committee of Guilin Medical University (Approval No: GLMC20221101). The approval covers collecting clinician behavioral data, including diagnosis decisions, time consumption, and all related operations, as well as the reconstruction and sharing of this data. Informed consent has been obtained from the clinicians involved. In this database, all personally identifiable information, except for the clinician's gender and age, has been either removed or regenerated in accordance with U.S. HIPAA regulations. 
Moreover, the patient data used in this database are derived from the publicly available MIMIC databases~\cite{johnson2016mimic,johnson2020mimic,johnson2019mimic,johnson2023mimic}. 
The authors have obtained permission to use this dataset, and no new ethical issues are involved. Additionally, the dataset is exclusively restricted for legitimate scientific research purposes. 


\vspace{0.2in}

\noindent \textbf{Data Collection and Processing.} AI.vs.Clinician represents a collaborative effort between Institute of Computing Technology, Chinese Academy of Science (ICT, CAS), International Open Benchmark Council (BenchCouncil), Guangxi Normal University, and fourteen critical medical centers like Guilin Medical University in China. Through this partnership, the human information and behavior data obtained from 125 clinicians in fourteen medical centers undergoes a rigorous process of deidentification and subsequent availability to qualified researchers who have made commitments to the lawful use of the data and not attempting to identify the individuals or institutes and disseminate the data.
AI.vs.Clinician has obtained the official ethical review from Guilin Medical University and the official approval to share the data under a series of commitments. In addition, the database has obtained informed consent from all the clinician participants. The population and demographics of clinician participants are shown in Table ~\ref{doctor-charac}, covering a broad spectrum of real-world population~\cite{zhan2024evaluatology}. Please note we only list the characteristics of 121 clinicians since the other four do not provide valid information.

Figure ~\ref{doctor} illustrates the creation steps of AI.vs.Clinician database, including (1) creation of a patient cohort, which will be diagnosed by clinicians and pre-trained AI models, and thus play pivotal roles in unraveling the interaction between the two entities, (2) data deduplication for AI training, (3) AI model training to the state-of-the-art or state-of-the-practice quality, (4) AI model inference on patient cohort, (5) clinician diagnosis on patient cohort with or without the assistance of AI models, (6) data validation, and (7) deidentification. 

\begin{table}[htbp]
\caption{Clinician Population and Demographics.}
\centering
\small
\definecolor{mycolor}{HTML}{EFAC8F}%
\begin{tabular}{|p{1.5in}|p{2.5in}|p{2in}|}
\hline
\rowcolor{mycolor}
\textbf{Category} & \textbf{Characteristics} & \textbf{No. of Clinicians (\% by unit)}   \\ \hline
\multirow{2}{2in}{\textbf{Gender}} & Male & 69 (57\%) \\ \cline{2-3}
& Female & 52 (43\%) \\ \hline
\multirow{4}{2in}{\textbf{Age}} & $age \leq 30$ & 30 (25\%) \\ \cline{2-3}
& $30 < age \leq 40$ & 49 (40\%) \\ \cline{2-3}
& $40 < age \leq 50$ & 35 (29\%) \\ \cline{2-3}
& $50 < age \leq 60$ & 7 (6\%) \\ \hline
\multirow{5}{2in}{\textbf{Years of Working}} & (0,5] & 31 (26\%) \\ \cline{2-3}
& (5,10] & 22 (18\%)  \\ \cline{2-3}
& (10,15] & 21 (17\%)  \\ \cline{2-3}
& (15,20] & 29 (24\%) \\ \cline{2-3}
& > 20 & 18 (15\%) \\ \hline
\multirow{4}{2in}{\textbf{Class of Position}} & None (During residency training) & 20 (17\%) \\ \cline{2-3}
& Junior (Resident physician) & 15 (12\%) \\ \cline{2-3}
& Intermediate (Attending physician)  & 41 (34\%) \\ \cline{2-3}
& Senior (Chief and Associate Chief Physician) & 45 (37\%) \\ \hline
\multirow{3}{2in}{\textbf{Institution Level}} & Grade-A Tertiary Hospital in China & 44 (36\%) \\ \cline{2-3}
& Grade-A Secondary Hospital in China & 76 (63\%) \\ \cline{2-3}
& Medical University & 1 (1\%) \\ \hline
\multirow{14}{2in}{\textbf{Department}} & Emergency Department & 32 (26\%) \\ \cline{2-3}
& Intensive Care Unit (ICU) & 44 (36\%)  \\ \cline{2-3}
& Internal Medicine & 16 (13\%) \\ \cline{2-3}
& Surgery Department & 6 (5\%) \\ \cline{2-3}
& Orthopedics & 1 (1\%) \\ \cline{2-3}
& Pediatrics & 5 (4\%) \\ \cline{2-3}
& Ophthalmology & 1 (1\%) \\ \cline{2-3}
& Gynaecology & 6 (5\%) \\ \cline{2-3}
& Traditional Chinese Medicine & 4 (3\%) \\ \cline{2-3}
& Gastroenterology & 1 (1\%) \\ \cline{2-3}
& Infectious Diseases Department & 1 (1\%) \\ \cline{2-3}
& Rheumatology and Immunology Department & 1 (1\%) \\ \cline{2-3}
& Neurology & 2 (2\%) \\ \cline{2-3}
& Anesthesia Department & 1 (1\%) \\ \hline
\end{tabular}
\label{doctor-charac}
\end{table}

We use open source and widely used clinical databases -- MIMIC (Medical Information Mart for Intensive Care) for sepsis detection and prediction, including MIMIC-III 1.4~\cite{johnson2016mimic}, MIMIC-IV 2.2~\cite{johnson2020mimic}, MIMIC-CXR-JPG 2.0.0~\cite{johnson2019mimic}, and MIMIC-IV-Note 2.2~\cite{johnson2023mimic}. Among them, 
The MIMIC-III contains over forty thousand patients who stayed in critical care units between 2001 and 2012~\cite{mimiciii}. The MIMIC-IV covers all medical records of patients admitted to an ICU or the emergency department between 2008 and  2019~\cite{mimiciv}, an updated version of MIMIC-III. MIMIC-CXR-JPG is an extensive publicly available database of labeled chest radiographs~\cite{johnson2019mimic}. MIMIC-IV-Note contains 331,794 de-identified discharge summaries from 145,915 patients and 2,321,355 de-identified radiology reports for 237,427 patients~\cite{mimicivnote}.


\subsection*{(1) Creation of Patient Cohort.}
We construct a patient cohort that contains a series of patient cases according to the criteria set by our collaborating clinicians, simultaneously ensuring demographic diversity.
Each patient case contains a patient ID (subject\_id in the dataset), a patient admission ID (hadm\_id in the dataset), and a current time frame. 
Note that the current time frame assumes the moment when a patient is being diagnosed, where both the clinician and the model can only access the data until the current time and have no visibility into future data.
It can encompass various stages for sepsis patients, including the pre-onset, onset, or post-onset periods of Sepsis, as well as an arbitrary time frame during hospital stays for non-sepsis patients.

\begin{figure}[h]
\centering
\includegraphics[scale=0.392]{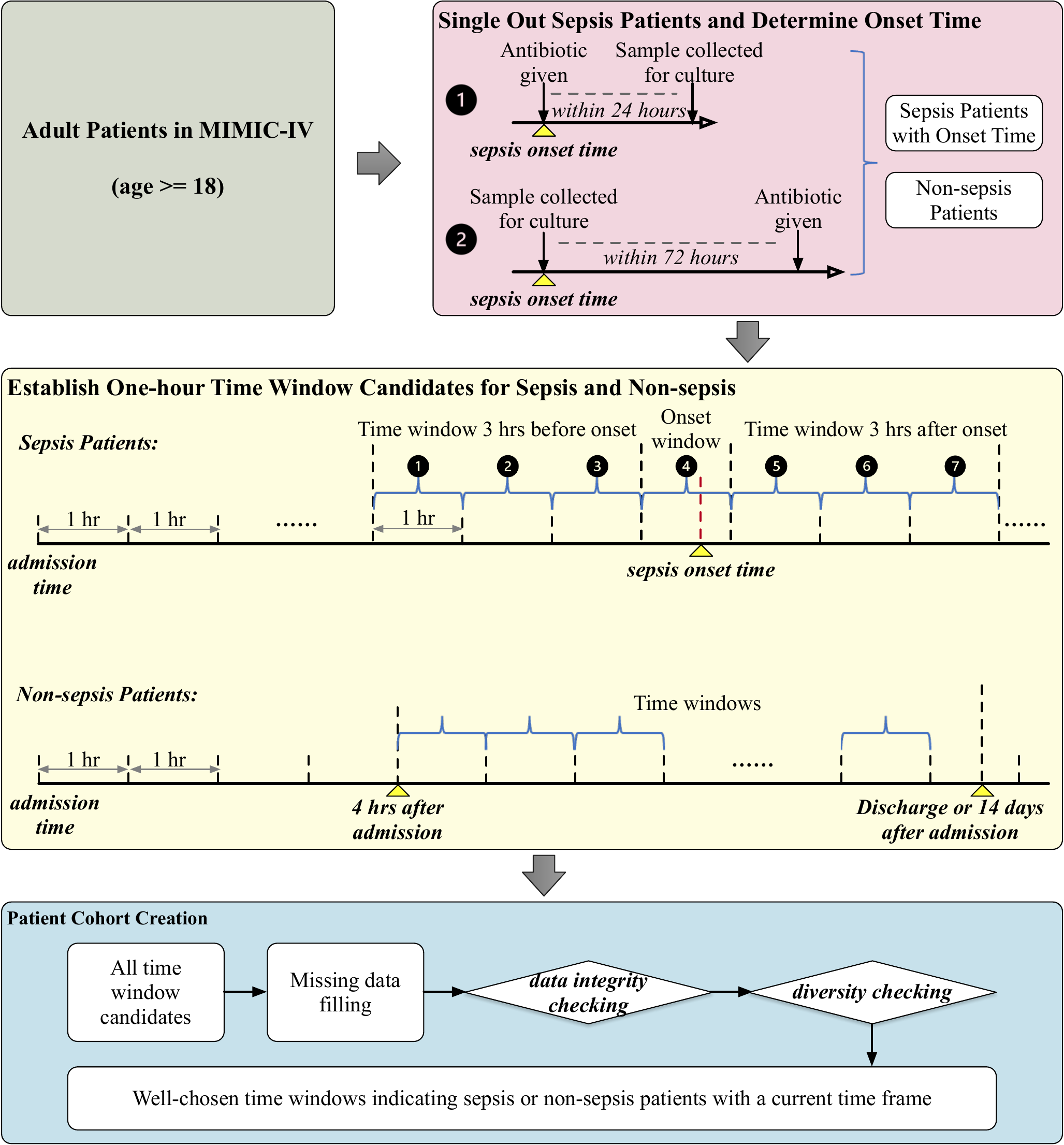}
\caption{Patient Cohort Creation.}
\label{cohort}
\end{figure}

We use MIMIC-IV, MIMIC-CXR-JPG, and MIMIC-IV-Note to choose 3000 patient cases containing medical imaging and comprehensive examination data. The patient cases contain 1500 positive cases (Sepsis) and 1500 negative ones (non-sepsis).
The choosing steps are as follows and shown in Figure ~\ref{cohort}.

\begin{itemize}
    \item For all adult ($age\geq18$) patients in MIMIC-IV dataset, we determine the onset time of Sepsis referring to the third international consensus definitions for Sepsis (Sepsis-3)~\cite{singer2016third,sepsischina} and researches published on Nature Medicine~\cite{komorowski2018artificial,sepsisonset} and JAMA~\cite{seymour2016assessment}. 
    During a patient's hospital stay, if he/she had qSOFA and SOFA scores~\cite{lambden2019sofa} greater than or equal to 2, we further check whether he/she was given antibiotics and collected samples for microbiological culture. If the antibiotic was given first and the microbiological sample was collected within 24 hours, then the given time of antibiotic was the onset time of Sepsis. If the sample for microbiological culture was collected first and the antibiotic was given within 72 hours, then the microbiological sample collection time is the onset time. The other patients with qSOFA and SOFA scores of less than two are considered patients without Sepsis (non-sepsis).
    
    \item For the patients with Sepsis, we start from their admission time and slide using a one-hour window until we cover the three hours after the onset of Sepsis. From a series of one-hour windows, we select time windows that fully cover the three hours, two hours, or one hour before the onset time of Sepsis. We also include the time window that contains the onset of Sepsis itself. Additionally, we choose time windows that entirely cover the one hour, two hours, or three hours after the onset of Sepsis. Thus, a patient can have a maximum of seven time windows been selected. Then, for the selected time windows, we further check whether the examination data within that period are complete and single out the time windows that contain all seven fundamental items and at least four advanced items as positive patient case candidates. Note that before the integrity check, we employed the missing data filling rules 
    referring to~\cite{jain2006decline,mao2018multicentre} for all the examination items, considering whether adverse events occurred and the validity period.
    The \textbf{seven fundamental items} are body temperature, systolic blood pressure, diastolic blood pressure, heart rate, respiratory rate, consciousness level, and 24-hour fluid input/output. The respiratory rate, consciousness level, and systolic blood pressure in fundamental items are used to calculate qSOFA score~\cite{singer2016third}.
    The \textbf{seven advanced items} are complete blood count, arterial blood gas analysis, hemostatic function, medical imaging data, pathogen screening, culture analysis, and smear tests.
    After that, we single out 8639 positive candidates.

    \item For each patient without Sepsis (non-sepsis), we start from four hours after admission to avoid leakage due to a short time range and iterate by the hour until the time of discharge or 14 days after admission. During this period, we identify and select all one-hour time windows that contain all seven fundamental items and at least four advanced items as negative patient case candidates. The total number is 3832.

    \item We choose 1500 positive cases and 1500 negative cases considering the demographic diversity, including age, gender, and weight. For positive cases, we also consider balancing the proportion of cases for the three-hour period before onset, the time of onset, and the three-hour period after onset. 
    Note that the total 3000 cases are from 2800 non-repetitive patients. For 100 out of the 2800 patients, we choose three time periods: before, during, and after the onset. For the other 2700 patients, we only include one time period that is either before, during, or after the onset.
    The 3000 cases are organized into five groups, each containing 300 positive cases and 300 negative cases. 
    Table ~\ref{patient-charac} shows the patient cohort population and demographics. 

\end{itemize}


\begin{table}[h]
\caption{Patient Population and Demographics of Patient Cohort.}
\centering
\small
\definecolor{mycolor}{HTML}{EFAC8F}%
\begin{tabular}{|p{1in}|p{1in}|p{0.58in}|p{0.58in}|p{0.58in}|p{0.58in}|p{0.58in}|p{0.58in}|}
\hline
\rowcolor{mycolor}
 &  & \multicolumn{6}{|c|}{\textbf{No. of Patients (\% by unit)}}   \\ \cline{3-8}
\noalign{\vskip-\arrayrulewidth}
\multirow{-2}{1in}{\cellcolor{mycolor}\textbf{Category}}  & \multirow{-2}{2in}{\cellcolor{mycolor}\textbf{Characteristics}}  & All & Group 1 & Group 2 & Group 3 & Group 4 & Group 5 \\ \hline
\multirow{2}{1in}{\textbf{Is sepsis?}} & Sepsis & 1500 (50\%) & 300  & 300 & 300 & 300 & 300 \\ \cline{2-8}
& Non-sepsis & 1500 (50\%) & 300 & 300 & 300 & 300 & 300 \\ \hline

\multirow{3}{1in}{\textbf{Current time for sepsis patients}} & At onset & 1000 (33\%) & 200 & 200 & 200 & 200 & 200 \\ \cline{2-8}
& 3 hours before onset & 250 (8\%) & 50 & 50 & 50 & 50 & 50 \\ \cline{2-8}
& 3 hours after onset & 250 (8\%) & 50 & 50 & 50 & 50 & 50 \\ \hline

\multirow{2}{1in}{\textbf{Gender}} & Male & 1704 (57\%) & 332 & 344 & 339 & 352 & 337 \\ \cline{2-8}
& Female & 1296 (43\%) & 268 & 256 & 261 & 248 & 263 \\ \hline

\multirow{7}{1in}{\textbf{Age}} & $18 \leq age \leq 30$ & 128 (4\%) & 22 & 30 & 31 & 26 &   19  \\ \cline{2-8}
& $31 \leq age \leq 40$ & 128 (4\%) & 18 & 25 & 24 &  31 &  30 \\ \cline{2-8}
& $41 \leq age \leq 50$ & 285 (10\%) & 55 & 65 & 65 &  53 &  47 \\ \cline{2-8}
& $51 \leq age \leq 60$ & 552 (18\%) & 106 & 111 & 113 &  116 &  106  \\ \cline{2-8}
& $61 \leq age \leq 70$ & 682 (23\%) & 144 & 125 &  145 & 127 &   141 \\ \cline{2-8}
& $71 \leq age \leq 80$ & 638 (21\%) & 147 & 122 & 119 &  124 &  126 \\ \cline{2-8}
& $age > 80$ & 587 (20\%) & 108 & 122  & 103 & 123 &  131 \\ \hline

\multirow{8}{1in}{\textbf{Weight (kg)}} & $Weight < 40 $ & 17 (1\%) & 2 & 5 & 3 & 4 & 3 \\ \cline{2-8}
& $41 \leq Weight \leq 50 $ & 110 (4\%) & 29 & 21 & 15 & 19 & 26 \\ \cline{2-8}
& $51 \leq Weight \leq 60 $ & 326 (11\%) & 72 & 80 & 54 & 63 & 57 \\ \cline{2-8}
& $61 \leq Weight \leq 70 $ & 549 (18\%) & 107 & 124 & 112 & 103 & 103 \\ \cline{2-8}
& $71 \leq Weight \leq 80 $ & 564 (19\%) & 114 & 109 & 115 & 114 & 112 \\ \cline{2-8}
& $81 \leq Weight \leq 90 $ & 508 (17\%) & 106 & 89 & 116 & 98 & 99 \\ \cline{2-8}
& $91 \leq Weight \leq 100 $ & 349 (11\%) & 62 & 76 & 67 & 76 & 68 \\ \cline{2-8}
& $ Weight > 100 $ & 577 (19\%) & 108 & 96 & 118 & 123 & 132 \\ \hline

\multirow{4}{1in}{\textbf{QSOFA Score}} & 0 & 756 (25\%) & 130 & 145 & 149 & 182 & 150 \\ \cline{2-8}
& 1 & 1435 (48\%) & 298 & 269 & 281 & 281 & 306 \\ \cline{2-8}
& 2 & 707 (24\%) & 142 & 165 & 157 &  117 & 126 \\ \cline{2-8}
& 3 & 102 (3\%) & 30 & 21 & 13 &  20 & 18 \\ \hline

\multirow{2}{1in}{\textbf{Mortality}} & Within 90 days for sepsis patients & \multirow{2}{0.58in}{13.27\%} & \multirow{2}{0.58in}{13.67\%} & \multirow{2}{0.58in}{15\%} & \multirow{2}{0.58in}{11.33\%} &  \multirow{2}{0.58in}{15.33\%} & \multirow{2}{0.58in}{11\%} \\ \hline

\end{tabular}
\label{patient-charac}
\end{table}




\subsection*{(2) Data Deduplication for AI Training.}

For data leakage prevention, we remove all patients selected as cases from the MIMIC datasets and divide the remaining patient admission IDs (hadm\_id) into training, validating, and testing sets for AI training. We first perform data deduplication by removing all the patient data that have already been chosen as patient cases from MIMIC-III and MIMIC-IV to use these two databases for joint training. After that, we obtain three sets, each containing a set of patient admission IDs that will be used as training, validating, or testing data in the following step, with a proportion of 7:1:2.
We establish 10528 non-repetitive patient admission IDs and partition them randomly into training, validating, and testing sets, with the numbers 7376, 1030, and 2122, respectively.
Detailedly,
the training set contains 3395 positive patient admission IDs (1732 from MIMIC-III and 1663 from MIMIC-IV) and 3981 negative patient admission IDs (676 from MIMIC-III and 3305 from MIMIC-IV).
The validating set contains 464 positive patient admission IDs (227 from MIMIC-III and 237 from MIMIC-IV) and 566 negative patient admission IDs (95 from MIMIC-III and 471 from MIMIC-IV).
The testing set contains 1012 positive patient admission IDs (537 from MIMIC-III and 475 from MIMIC-IV) and 1110 negative patient admission IDs (167 from MIMIC-III and 943 from MIMIC-IV).



\subsection*{(3) AI Model Training.}

We further investigate the state-of-the-art or state-of-the-practice AI algorithms for sepsis detection and prediction to explore the interaction between AI and clinicians.
Cox proportional hazards model (CoxPHM)~\cite{henry2015targeted,henry2022factors,adams2022prospective} and long short-term memory (LSTM)~\cite{kaji2019attention} are representative models from survival analysis and deep learning~\cite{cohen2024subtle}, respectively, and are the two most popular and widely used AI algorithms for sepsis detection and prediction. We choose features from the patient's examination data according to ~\cite{henry2015targeted,henry2022factors,adams2022prospective} for a CoxPHM model and ~\cite{kaji2019attention} for an LSTM model, respectively, and further implement detection versions and prediction versions of these two models with different model qualities. Among them, the detection version outputs the probability of the sepsis onset at present, and we use 0h to represent this type. The prediction version outputs the probability of the sepsis onset within the next three hours, and we use 3h to represent this type. The detection and prediction versions for two models with AUC (area under the curve) 0. XX is called LSTM-0h-AUCXX, and LSTM-3h-AUCXX for short. For example, LSTM-0h-AUC95 means training an LSTM model with AUC 0.95 to detect whether a patient is in sepsis onset. CoxPHM-0h and CoxPHM-3h are two models with AUC 0.95.
The details are shown in Fig.~\ref{trainsepsis}.

\begin{itemize}

    \item \emph{LSTM Model Features and Labels.} For LSTM models, We organize 336 (14 days * 24 hours/day) lines of patient examination data as a training, validating, or testing sample. Starting from the patient's admission time, each line contains the examination data within one hour until the end condition or reaching 14 days. If the number of lines is less than 336, the remaining lines are filled with the value of -4. 
    For LSTM-0h, the positive samples consist of two parts: patients who are in the sepsis onset phase and patients who have been experiencing Sepsis for three hours. For the former part, we start from their admission time and slide using a one-hour window until we cover the onset time of Sepsis.
    For the latter part, we also start from their admission time and slide using a one-hour window until we cover the three hours after the onset of Sepsis. 
    When there are missing examination items, priority is given to supplementing them with adjacent data from the same patient within the valid period, provided that no adverse events have occurred. If this condition is not met, the missing items are supplemented with the average values of the corresponding data items from all patients at the same time interval from the onset of Sepsis.
    The negative samples also contain two parts: non-sepsis patients and sepsis patients more than 12 hours away from the onset time. Since non-sepsis patients have no onset time, the end condition is based on a Gaussian distribution of the length of the hospital stay for sepsis patients to avoid model leakage. 
    For LSTM-3h, the positive and negative samples adopt the same strategy as LSTM-0h. The difference is that the LSTM-3h adds the patients who will experience sepsis onset within the next three hours as the positive samples.


    \item \emph{CoxPHM Model Features and Labels.} Contrary to the LSTM models that start from the admission time, the CoxPHM models start from the sepsis onset time and slide backward toward the admission time using a one-hour window. 
    \item \emph{Hyperparameter Settings.} The hyperparameter settings are as follows.
    LSTM-0h-AUC95 and LSTM-3h-AUC95: using batchsize 32, learning rate 0.0001, and convergence training. LSTM-0h-AUC85: using batchsize 32, learning rate 0.003, and training only three epochs. LSTM-0h-AUC75: using batchsize 32, learning rate 0.007, and training only two epochs. LSTM-3h-AUC85: using batchsize 32, learning rate 0.005, and training three epochs. LSTM-3h-AUC75: using batchsize 32, learning rate 0.00715, and training two epochs. CoxPHM-0h: l1\_ratio 1, penalizer 0.05, step\_size 0.1, precision 1e-08, and max\_steps 1000.
    \item \emph{Model Quality.} 
Our trained models have achieved state-of-the-art model quality comparable to the results reported in relevant research. The AUC on testing data for LSTM achieves 0.95. In addition, in order to explore the impact of model quality on human decision, we further train two LSTM models with a medium quality (AUC85) and a low quality (AUC75). Specifically, the AUCs for LSTM-0h-AUC95, LSTM-0h-AUC85, LSTM-0h-AUC75, LSTM-3h-AUC95, LSTM-3h-AUC85, and LSTM-3h-AUC75 are 0.9468, 0.8497, 0.7447, 0.9565, 0.8506, and 0.7565, respectively. The AUCs on testing data for CoxPHM-0h and CoxPHM-3h are 0.95 and 0.92, respectively.

\begin{figure}[h]
\centering
\includegraphics[scale=0.7]{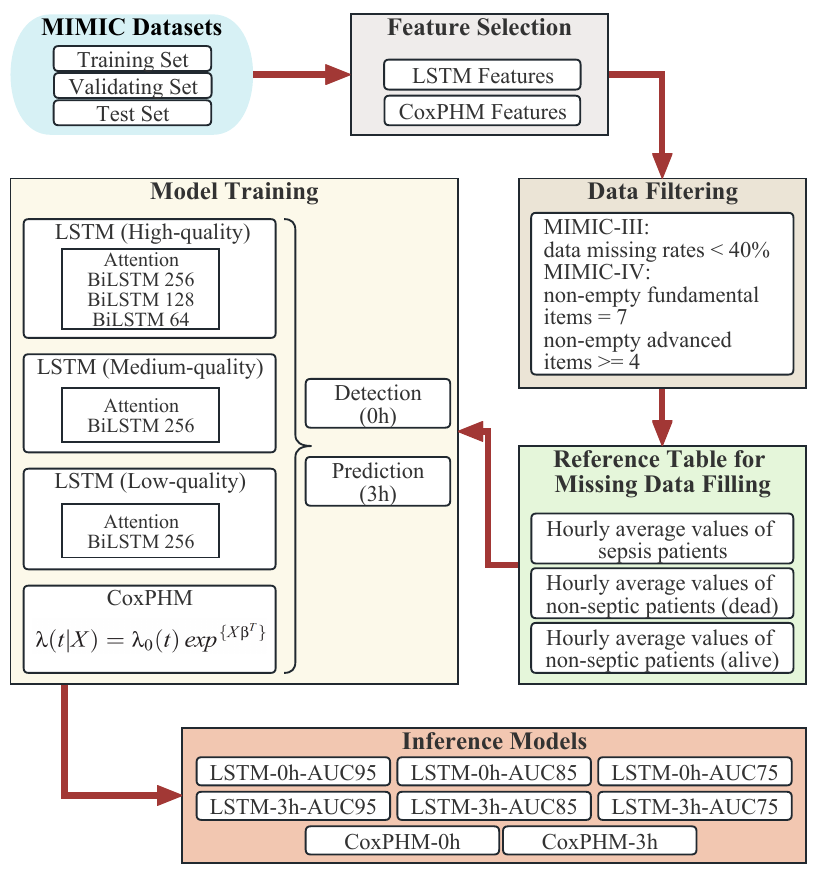}
\caption{The Detailed Training Process for LSTM and CoxPHM.}
\label{trainsepsis}
\end{figure}

\end{itemize}

\subsection*{(4) AI Model Inference on Patient Cohort.}

We use the trained models to predict every patient case within the well-chosen patient cohort. We collect the corresponding features required by the models for all the patient cases. Note that the end time is the current time frame specified in each patient case to ensure the latter information is inaccessible for both the clinicians and AI models.
The model quality of the patient cases is as follows: The AUCs for LSTM-0h-AUC95, LSTM-0h-AUC85, LSTM-0h-AUC75, LSTM-3h-AUC95, LSTM-3h-AUC85, and LSTM-3h-AUC75 are 0.9974, 0.9808, 0.8088, 0.9999, 0.9854, and 0.8325, respectively. The AUCs for CoxPHM-0h and CoxPHM-3h are 0.92 and 0.98, respectively.

\subsection*{(5) Clinical Trials using AI Models.}

In this step, we perform clinical trials using AI models in our collaborative medical centers. The clinicians will diagnose the patient cases within the well-chosen patient cohort with or without the assistance of different AI models. 
We build a specialized early warning system for clinicians' diagnoses and treatments, recording all their operations and timestamps, including the login, logout, diagnosis, treatment, clicked examination items, and all corresponding timestamps. 
Note that for each patient case, the clinician first performs preliminary diagnosis according to the current fundamental examination data and all the history examination data with or without the probability output of AI models. After the preliminary decision, the clinician can click on the current advanced examination data for more information and make a final decision. The clinicians can choose from five options, severe Sepsis, Sepsis, high suspicion, low suspicion, and non-sepsis, to make a preliminary or final decision, or they can enter the text in the input box directly. In addition, they can provide treatment and drug regimens through a textbox after the preliminary and final diagnosis.
All the human-AI interaction data are recorded in this specialized system.

\vspace{0.1in}

\noindent The process of clinician recruitment is as follows:

Step 1: Early warning system on Sepsis implementation, integrating four AI models and a random model.

Step 2: Promotion in multiple medical centers in Guangxi province, China.

Step 3: Conduction of two online training sessions on how to use the early warning system.

Step 4: Clinician recruitment and selection. The inclusion criteria for clinicians are: 

\setlength{\parindent}{5em} (a) possession of a medical license;

(b) having clinical experience in diagnosing and treating sepsis.

\setlength{\parindent}{1.5em} Step 5: Early warning system deployment in medical centers where clinicians were recruited.

Step 6: Data collection and analysis.

\subsection*{(6) Data Validation.}

We validate and single out effective data according to the following criteria:

\begin{itemize}
    \item The clinician's information is complete.
    \item The diagnosis processes are complete, including both the preliminary and final stages.
    \item The timestamp of the final diagnosis is more than that of the preliminary stage.
    \item At least one diagnosis conclusion of the preliminary or final stage explicitly indicates a sepsis-related diagnostic decision.
\end{itemize}

\subsection*{(7) Deidentification.}

Following the stipulations and identifiers defined by The Health Insurance Portability and Accountability Act (HIPAA), we remove all the sensitive identifiers to de-identify the clinician's information, including the name, location, national identification number, phone number, email address, and institution. We also randomize the ID number in our database so that the clinicians from the same institution will be shuffled. 
Additionally, for each clinician, timestamps are shifted into the future with a random offset, and this shift is done consistently to ensure that the behavior intervals within the same clinician remain preserved.





 

\section*{Data Records}


We adopt a consistent or similar scheme for our database construction to be compatible with MIMIC databases. AI.vs.Clinician adopts a relational type and contains 22 tables. Partial identifiers used for the tables are the same as the MIMIC-III and MIMIC-IV databases, for example, subject\_id and hadm\_id, so that the users can directly locate the corresponding patient in the MIMIC databases.

AI.vs.Clinician provides the patient related, AI model related, and clinician related tables. In terms of the patient cohort, the database provides 14 tables, including the patient case information, the current and history fundamental examination items, the current and history advanced examination items, SOFA, drug history, medical history, and chief complaint. These data are either reorganized or analyzed based on the MIMIC databases and the creation criteria of the patient cohort.
Regarding AI models, the database provides five tables, including the model properties like Model\_ID, Model\_Name, sensitivity, specificity, precision, AUC (area under the curve), etc., the model datasets used for training, validating, and testing, the input features used by the models, and the inference results on the patient cohort.
With respect to clinicians, the database provides three tables, including the clinician's information like Clinician\_ID, Insitution\_Level, gender, etc., the clinician's click behaviors like clicking and viewing an examination item, and the clinician's preliminary and final diagnosis decisions and treatment on a patient case with or without the aid of an AI model. 
The details about the 22 tables are listed in Table ~\ref{haimed-tables}.

\begin{table}[h]
\caption{Overview of Tables in AI.vs.Clinician Database.}
\centering
\small
\definecolor{mycolor}{HTML}{EFAC8F}%
\begin{tabular}{|p{1.8in}|p{4.6in}|}
\hline
\rowcolor{mycolor}
\textbf{Table Name} & \textbf{Description} \\ \hline
Patient\_Case\_Info & The information for each patient case, including the identifier information, demographics, admission time, Sepsis or non-sepsis, sepsis onset time, and current time window. \\ \hline
Patient\_Fundamental\_Item & The current and history information of seven fundamental examination items for patient cases. \\ \hline
Fluid\_Input & The latest 24-hour and history fluid input information for each patient case. \\ \hline
Fluid\_Output & The latest 24-hour and history fluid output information for each patient case. \\ \hline
Complete\_Blood\_Count & The current and history complete blood count information for each patient case. \\ \hline
Pathogen\_Blood & The current and history pathogen blood information for each patient case. \\ \hline
Medical\_Imaging & The current and history medical imaging information for each patient case. \\ \hline
Arterial\_Blood\_Gas\_Analysis & The current and history arterial blood gas analysis data for each patient case. \\ \hline
Hemostatic Function & The current and history hemostatic data for each patient case. \\ \hline
Culture\_Smear & The current and history culture smear information for each patient case. \\ \hline
Procalcitonin & The current and history procalcitonin information for each patient case. \\ \hline
SOFA & The current variables of six types related to the computation of SOFA score, referring to~\cite{singer2016third,sepsischina,seymour2016assessment} for each patient case. \\ \hline
Drug\_History & The drug history information for each patient case. \\ \hline
MedicalHistory\_ChiefComplaint & The medical history and chief complaint information for each patient case.\\ \hline
Model\_Property & The properties of each AI model, including the model ID, model name, the sensitivity, specificity, precision, and AUC on training, validating, and testing dataset.\\ \hline
Model\_Dataset & The patient admission IDs (hadm\_id) used as the training set, validating set, or testing set for all the AI models. \\ \hline
CoxPHM\_Feature & The feature input for each patient case to train a CoxPHM model.  \\ \hline
LSTM\_Feature &  The feature input for each patient case to train an LSTM model. \\ \hline
Model\_Cohort\_Infer & The inference results of four AI models and one random model on the patient cohort, including the decision, the probability of sepsis onset presently (0h), and the probability of sepsis onset within the next three hours (3h).  \\ \hline
Clinician\_Info & The information for each clinician, including the identifier information, demographics, department, years of working, class of position, and area of expertise.  \\ \hline
Clinician\_Click\_Behavior & The information of clinician's click behaviors, including the clicked examination item for viewing, and the clicked time. \\ \hline
Clinician\_Diagnosis\_Treatment & Each clinician's preliminary and final diagnosis decisions and treatment on each patient case with or without the aid of an AI model, including the clicked sequence (link with Clinician\_Click\_Behavior table), preliminary decision, final decision, preliminary treatment, final treatment, and corresponding timestamps. \\ \hline

\end{tabular}
\label{haimed-tables}
\end{table}

\section*{Technical Validation}


The construction process of the AI.vs.Clinician database underwent a series of procedural and manual validations and assessments, rigorously addressing aspects such as correctness, integrity, consistency, deidentification, and ethics. 
Version control software was utilized for code management, while all the data processing or transformation operations were performed using a reproducible script.

In processing patient data, we strictly adhere to the requirements and specifications of the MIMIC database, conducting thorough data verification and validation. For instance, when using MIMIC-III and MIMIC-IV, as their periods overlap and may contain duplicate patients, we rigorously screen and deduplicate the data to prevent issues of data leakage during AI model training. We select state-of-the-art or state-of-the-practice algorithms for AI models and ensure that the model achieves comparable performance to the reference papers. In the diagnostic phase with clinicians, we construct an early warning system and confirm the correctness and completeness of the data through extensive discussions with numerous clinicians. We strictly adhere to the rules of clinical trials in selecting participating clinicians, setting control groups, and conducting multiple rounds of system usage training. With consent, all actions are recorded by the early warning system. We verify the completeness and correctness of the data through procedures and manually, such as excluding data where diagnoses are incomplete or irrelevant, 
to ensure the rationality of the data.

Based on all the records and logs from the early warning system, the AI.vs.Clinician database is constructed through a series of processes, including integrity checks, consistency checks, and deidentification checks. Specifically, we create a set of rules and unit tests to check all the tables within the database.


\section*{Usage Notes}

\textbf{Data Access.} AI.vs.Clinician database comprises a set of comma-separated value (CSV) files and all related source code. Since the database contains not only the patients' information from MIMIC databases but also the clinicians' information from 14  medical centers, users must use the database with caution and respect. 
The database has been uploaded to PhysioNet~\cite{goldberger2000physiobank} platform (waiting for approval) under the ``Contributor Review" access policy. We also upload the database to the Journal's online submission system for review.
To access the database, the following steps need to be completed:
\begin{itemize}
    \item The researchers must complete the access steps required by MIMIC databases.
    \item The researchers are required to sign a data use agreement, which delineates acceptable data usage and security protocols and prohibits attempts to identify individual clinicians and patients.
    \item The researchers are required to send an access request to the contributors and provide a description of the research project.
\end{itemize}

\noindent \textbf{Example Usage.} AI.vs.Clinician provides the first database for the research on human and AI interaction, which holds significant importance for developing AI in medicine and translating AI into clinical practice. AI.vs.Clinician facilitates a wide array of research investigations, such as AI algorithm benchmarking, optimization, and human-in-the-loop research in medicine. 

AI.vs.Clinician provides a collection of source codes used during its construction, e.g., the creation of patient cohort, data deduplication, AI model training and inference, data validation, etc., as well as the related code for processing and analysis. 
The instructions for using the database are illustrated in a README file in the database package. We will further expand the dataset and provide more comprehensive scripts. 



\section*{Code availability}

Since AI.vs.Clinician database is based on MIMIC database, the users are required to be a credentialed user of MIMIC and complete the required training. 
Due to the sensitive clinician information, the full diagnosis records and logs cannot be publicly released. However, besides the reorganized database illustrated in this paper, we also provide an original version, which has been performed in the same process of validation and deidentification. The differences are that the original version uses the original table structures from the early-warning system and provides both English and Chinese information.
After identification, the database is publicly available from PhysioNet (awaiting approval). We also uploaded the database to the journal's online submission system for review. All the source code for creating patient cohort,  AI training and inference, etc., is available from \url{https://github.com/BenchCouncil/AI.vs.Clinician}. 

\bibliography{sample}


\section*{Acknowledgements}


We acknowledge supports from the Innovation Funding of ICT, CAS under Grant No. E461070. We thank all the participating medical centers and clinicians.

\section*{Author contributions statement}


W.G. conceptualized this study, conceived the experiments, implemented the CoxPHM model according to the referenced paper, and wrote the manuscript. Y.L. conceptualized this study and defined the patient cohort criteria. Z.Y., D.C., W.L., X,L., J.Z., J.X., and H.L. implemented the LSTM models and early warning system and collected and analyzed the data. L.M., N.Y., Y.K., D.L., P.P., W.H., Z.L., J.H., F.H., G.Z., C.J., and T.W. recruited clinicians and participated in the experiment. Z.Z., Y.H., and J.Z. conceptualized this study, directed the project, and revised the manuscript. All authors have read and approved the final manuscript.

\section*{Competing interests} 

The authors declare no competing interests.

\end{document}